\pdfoutput=1

\documentclass[11pt]{article} 
\usepackage{acl}

\usepackage{times}
\usepackage{latexsym}

\usepackage[T1]{fontenc}
\usepackage[utf8]{inputenc}
\usepackage{microtype}
\usepackage{inconsolata}
\usepackage{graphicx}

\usepackage{soul}
\setul{0.5ex}{0.3ex}  
\setulcolor{red}      
\usepackage[normalem]{ulem}
\usepackage{pifont}
\usepackage{amssymb}
\usepackage{amsmath}
\usepackage{float}
\usepackage{hyperref}
\usepackage{url}
\usepackage{booktabs}
\usepackage{wrapfig}
\usepackage[dvipsnames]{xcolor}
\usepackage{enumitem}
\usepackage{lineno}
\usepackage{array}
\usepackage{booktabs}
\usepackage{caption} 
\usepackage{enumitem}
\usepackage{placeins}

\setlist[itemize]{noitemsep, topsep=0pt}

\usepackage{tcolorbox}
\tcbuselibrary{listings,skins,breakable}

\newtcblisting{promptbox}[1][]{
    enhanced,
    breakable,
    attach boxed title to top center={yshift=-3mm,yshifttext=-1mm}, 
    colback=white,
    colframe=black,
    colbacktitle=black!75!yellow,
    boxed title style={size=small},
    listing only,
    listing options={
        breaklines=true,        
        breakatwhitespace=true, 
        basicstyle=\small\ttfamily,
        breakindent=0pt
    },
    #1
}

\lstdefinestyle{boldwords}{
  basicstyle=\scriptsize\ttfamily,
  moredelim=[is][\bfseries]{**}{**} 
}

\newtcblisting{mainpromptbox}[1][]{
    enhanced,
    breakable,
    width=0.49\textwidth,
    attach boxed title to top center={yshift=-3mm,yshifttext=-1mm}, 
    colback=white,
    colframe=black,
    colbacktitle=black!75!yellow,
    boxed title style={size=small},
    listing only,
    listing options={
        breaklines=true,
        breakatwhitespace=true,
        breakindent=0pt,
        style=boldwords
    },
    #1
}

\colorlet{colexam}{brown!50!black}
\tcbset{
  base/.style={
    empty,
    frame engine=path,
    colframe=brown!10,
    sharp corners,
    title={Generated Rationale},
    attach boxed title to top left={yshift*=-\tcboxedtitleheight},
    boxed title style={size=minimal, top=4pt, left=4pt},
    coltitle=colexam,
    fonttitle=\small\bfseries\sffamily,
  }
}
\newtcolorbox[use counter=example]{myexamplea}{%
  base,
  colback=colexam,  
  boxed title style={
    overlay={
      \draw[colexam,line width=2pt] (frame.north west)--(frame.north east);
    }
  },
  overlay unbroken={
    \draw[colexam] ([yshift=-1.5pt]title.north east)--([xshift=-0.5pt, yshift=-1.5pt]title.north-|frame.east);
  }
}

\colorlet{colnovelty}{green!70!black}

\newtcolorbox[use counter=figure]{noveltyassessment}[3]{%
  empty,
  frame engine=path,
  colframe=yellow!10,
  sharp corners,
  title={#1},
  attach boxed title to top left={yshift*=-\tcboxedtitleheight},
  boxed title style={
    size=minimal, 
    top=4pt, 
    left=4pt,
    overlay={\draw[colexam,line width=2pt,] (frame.north west)--(frame.north east);}
  },
  coltitle=colexam,
  fonttitle=\small\bfseries\sffamily,
  colback=colexam,
  overlay unbroken={
    \draw[colexam] ([yshift=-1.5pt]title.north east)--([xshift=-0.5pt, yshift=-1.5pt]title.north-|frame.east);
  },
  after={\captionof{figure}{#2}\label{#3}\medskip}
}

\definecolor{darkblue}{rgb}{0, 0, 0.5}
\hypersetup{colorlinks=true, citecolor=darkblue, linkcolor=darkblue, urlcolor=darkblue}

\title{GraphMind: Interactive Novelty Assessment System \\ for Accelerating Scientific
Discovery}

\author{Italo Luis da Silva, Hanqi Yan, Lin Gui,  Yulan He \\
Department of Informatics\\
King's College London, UK \\
\texttt{\{italo.da\_silva,hanqi.yan,lin.1.gui,yulan.he\}@kcl.ac.uk} \\
}
\begin{document}

\maketitle

\begin{abstract}

Large Language Models (LLMs) show strong 
reasoning and text generation capabilities, prompting their use in scientific literature analysis, including novelty assessment.
While evaluating novelty of scientific papers is crucial for peer review, it requires extensive knowledge of related work, something not all reviewers have. 
While recent work on LLM-assisted
scientific literature analysis supports literature comparison, existing
approaches offer limited transparency and lack mechanisms for result
traceability via an information retrieval module. To address this gap, we
introduce \textbf{GraphMind}, an easy-to-use interactive web tool designed to
assist users in evaluating the novelty of scientific papers or drafted
ideas. Specially, \textbf{GraphMind} enables users to capture the main
structure of a scientific paper, explore related ideas through various
perspectives, and assess novelty via providing verifiable contextual insights.
\textbf{GraphMind} enables users to annotate 
related papers through various relationships,  and assess novelty with
contextual insight. This tool integrates 
Semantic Scholar with LLMs to support annotation, 
classification of papers. This combination provides users with a rich,
structured view of a scientific idea's core contributions and its connections to
existing work. 
\textbf{GraphMind} is available at
\url{https://oyarsa.github.io/graphmind} and a demonstration video at \url{https://youtu.be/wKbjQpSvwJg.} 
The source code is available at \url{https://github.com/oyarsa/graphmind}.
    
\end{abstract}
\section{Introduction}
 
Peer reviewing of scientific papers is a challenging task essential for
scientific collaboration. It requires a thorough understanding of the paper
being evaluated, as well as knowledge of the scientific literature around the topic. However, with the increasing number of publications, ranging from peer-reviewed conference and journal articles to preprints,  it is increasingly
difficult for reviewers to stay up to date within their research domains.
Concurrently, advances in LLMs’ reasoning and text generation capabilities have
spurred interest in their use for scientific literature review~\citep{yuan2021automatescientificreviewing,LIN2023AutoScholarReview,chitale2025autorev}.

There are two key dimensions to consider when evaluating a research paper. 1) at the \textbf{macro level}, how does the paper relate to existing work? Has similar research already been published? Compared to the cited or relevant prior work, does this paper present a groundbreaking idea or merely an incremental contribution? 2) at the \textbf{micro level}, is the paper well-organized and clearly motivated? How is the structure laid out, and do the sections logically support one another? Is there coherence across chapters, and does each part contribute meaningfully to the overall argument? 3) Building on these two dimensions, a third consideration is the \textbf{interaction between macro and micro levels}: can each micro-level component (e.g., specific methods, results, or arguments) be supported or contextualized by the macro-level literature? In other words, does the paper consistently draw connections between its internal structure and the broader research landscape? 4) Finally, based on all of the above, the ultimate question is: how can we synthesize these observations into an evaluation metric that effectively measures the paper’s overall contribution?

However, existing review tools often fall short in capturing all the key
features necessary for comprehensive paper evaluation and remain limited in
scope. Some focus primarily on retrieving potentially related papers via
citation networks, such as Connected
Papers\footnote{\url{https://connectedpapers.com/}} and
Inciteful\footnote{\url{https://inciteful.xyz/}}, while others rely on
surface-level semantic search, such as keyword or title matching, as seen in
tools like Litmaps\footnote{\url{https://litmaps.com/}}. Additional systems
attempt to assess paper quality based on content
alone~\citep{lin2024evaluatingenhancinglargelanguage,kang2018datasetpeerreviewspeerread}.
In general, these tools typically lack integration between the internal content
of a paper and its broader research context, which is an essential factor for
evaluating scientific novelty.

To address these limitations, we propose \textbf{GraphMind}, an interactive tool designed to support novelty assessment by analyzing both macro-level and micro-level information, as well as the interaction between them. \textbf{GraphMind} enables a systematic and comprehensive analysis of a paper’s contribution. In practice, it allows users to identify key research components and explore related work through a combination of citation-based and semantic relationships. Unlike traditional methods, our system decomposes abstracts into distinct background and target components, which enhances its ability to retrieve semantically related papers. This deeper contextual understanding facilitates a more robust and informed evaluation of novelty.

To support this functionality, we leverage \textbf{a combination of existing tools to collect macro-level information}. For instance, using the arXiv API, we can search for papers and retrieve their full LaTeX content. Additionally, the Semantic Scholar API\footnote{\url{https://www.semanticscholar.org/product/api}} allows us to access metadata, citation information, and recommended related papers.
At the \textbf{micro level, we utilize LLMs}, such as GPT-4o\footnote{\url{https://openai.com/index/gpt-4o-system-card/}} and Gemini 2.0 Flash\footnote{\url{https://deepmind.google/technologies/gemini/}}, to extract key elements from each paper, including claims, methodologies, experiments, background, and research objectives. Using this information, we construct a hierarchical graph of related papers, integrating both top-cited works and semantically similar papers, to support novelty assessments. This graph forms the basis for generating reports that highlight both supporting and contrasting evidence from the literature.

To ensure flexibility and usability, GraphMind features a web-based frontend
that allows users to explore and select papers for evaluation, and a backend
server that handles API queries and vector-based similarity search. Users can
operate in three modes: (1) browsing a curated set of papers with pre-computed
analyses, (2) dynamically evaluating new papers from arXiv using live data
retrieval and analysis, or (3) evaluating draft papers by directly entering the
title and abstract:

\begin{itemize}
    \item We introduce \textbf{GraphMind}, a tool designed to assist paper reviews by generating structured reports that combine a paper's key elements with insights from related works.
    \item We leverage APIs from arXiv and Semantic Scholar, with LLM-based extraction, to identify the key aspects of scientific papers.
    \item Our tool allows the creation novelty assessment reports that integrate detailed analysis of a paper's contribution and its position within the surrounding research landscape.
\end{itemize}
\section{Related Work}

There has been substantial progress in both macro and micro-level for LLM-assisted relevant paper recommendations and automated peer review. 

\paragraph{Macro level: Automated paper recommendation.} Existing retrieval
approaches focus on citation-based connections rather than content-level
analysis \cite{doi:10.1126/science.1240474,Beyond_surface_correlations2025,kreutz2022scientificpaperrecommendationsystems}. Tools like Connected Papers, Inciteful,
Litmaps, and
ResearchRabbit\footnote{\url{https://researchrabbitapp.com}} build paper graphs
using co-citation, bibliographic coupling, or title/keyword similarity, but
often miss deeper semantic relationships and lack transparency in their
algorithms. SPECTER~\citep{Singh2022SciRepEvalAM} improves relatedness scoring
via citation-informed embeddings, but does not extract or leverage specific
content from papers. Scholar Inbox \cite{flicke2025scholarinboxpersonalizedpaper} considered both citation and semantic information, but only document level without fine-grained analysis.
\citet{Guo_Chen_Zhang_Liu_Dong_He_2020} use title-abstract attention relations for retrieval,
but do not identify detailed semantic relationships between components.

\paragraph{Micro level: LLM-assisted scientific paper review.} Other works focus on evaluating and reviewing papers exclusively from their own content.
PeerRead~\citep{kang2018datasetpeerreviewspeerread} includes a small
subset with expert-annotated aspects such as clarity, impact, and originality.
SciND~\citep{article} constructs a knowledge graph from extracted novel entity
triplets in publications to support novelty assessment, but it does not provide
direct novelty annotations.
SchNovel~\citep{lin2024evaluatingenhancinglargelanguage} extracted abstracts and
metadata (e.g., institution, publication year) from 150,000 papers in the arXiv
dataset\footnote{\href{https://www.kaggle.com/datasets/Cornell-University/arxiv}{https://www.kaggle.com/datasets/Cornell-University/arxiv}}.
However, it infers novelty through publication date, assuming newer work is more novel, a simplistic proxy that overlooks semantic content. 
\citet{ai2024novascorenewautomatedmetric} determines the novelty of a document
by comparing its atomic content units (ACUs). It retrieves similar ACUs by
cosine similarity and calculates the novelty depending on how salient the ACUs
are in the corpus. While algorithmically interesting, this approach fails to
capture scientific innovation and creativity.

In summary, existing tools either lack macro or micro-level understanding, barely discuss the information interaction, such as supporting evidence between two levels. Therefore, GraphMind aims to fill in the gap by combining structured paper understanding with relationship-aware analysis, to further support the novelty assessment.

\section{Architecture of GraphMind}

\begin{figure}
    \centering
    \includegraphics[width=\columnwidth]{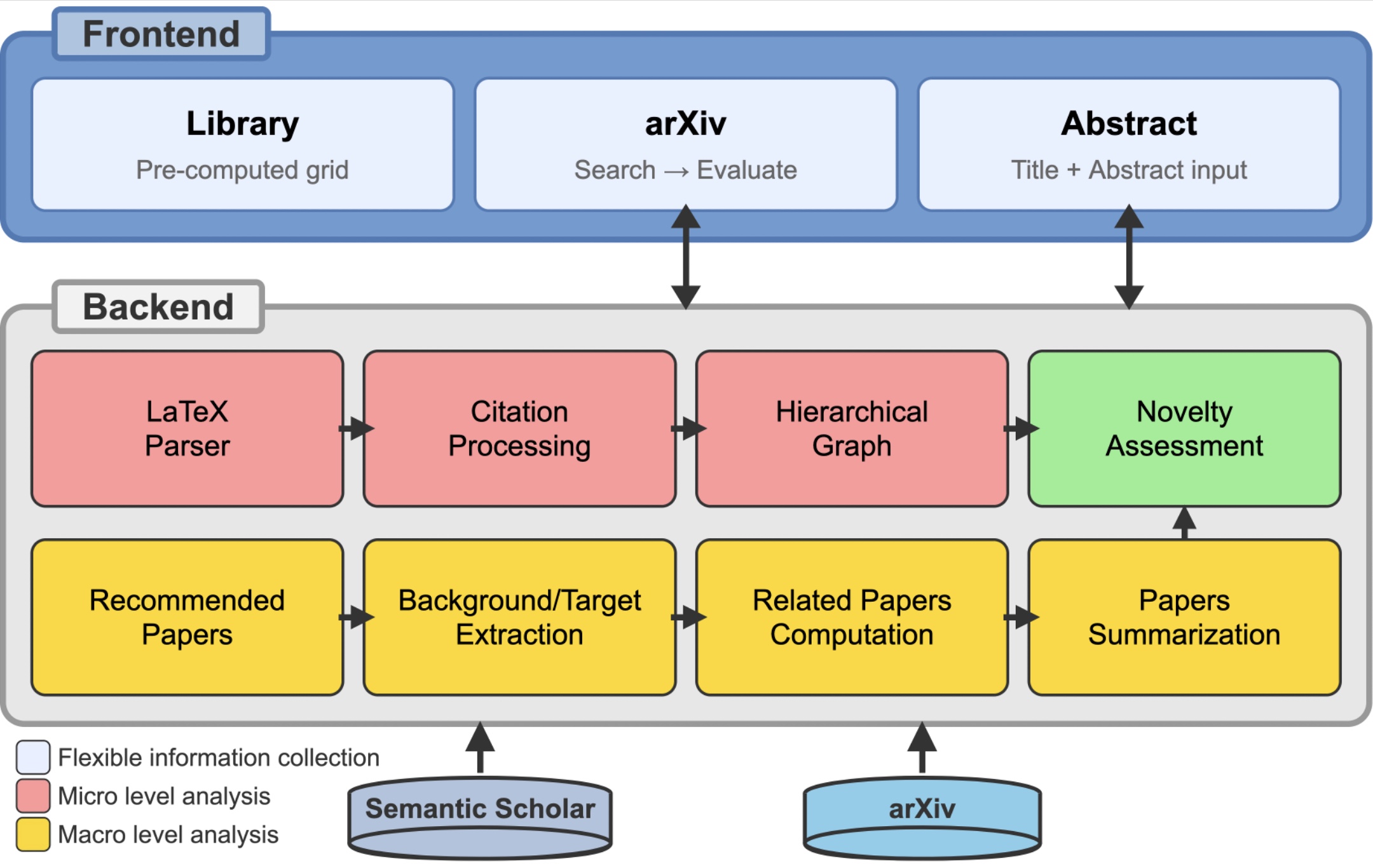}
    \caption{The overall architecture of \textbf{GraphMind}.}
    \label{fig:tool-architecture}
\end{figure}

\textbf{GraphMind} is composed of a web-based frontend and a backend server. Together, they enable users to search, analyze, and evaluate the novelty of scientific papers. An overview of the system is presented in Figure~\ref{fig:tool-architecture}. The \textbf{Frontend} is designed to support a more flexible information collection paradigm, while the \textbf{Backend} focuses on enabling a comprehensive analysis of the given paper by considering both macro- and micro-level information.

\paragraph{Frontend} 
The web interface is built using vanilla TypeScript as a Multi-Page Application, with separate files for each
page. It communicates with the server via a REST API. 
There are two main pages: \textit{Search} and \textit{Detail}.
In the \textit{Search} page, the user can select papers from either a pre-computed \textit{Library} or perform a dynamic \textit{arXiv}
search. 
The \textit{Library} contains a selected subset of papers from the ICLR 2022-2025
and NeurIPS 2022-2024 conferences, that have been fully pre-processed. This enables quick access without relying on external APIs. 
The \textit{arXiv} search allows users to query any arXiv paper. If data is available, the system runs the full evaluation pipeline. A progress bar is shown during processing, and users can enable notifications upon completion. Results are cached locally to avoid repeated processing. 
Regardless of how the user chose the paper, they eventually arrive at the \textit{Detail}
page, where they can see key metadata for the paper, extracted information, related works, 
and the final structured novelty report.

\paragraph{Backend} 

The server is written in Python with FastAPI. It presents a
REST API with three endpoints: \texttt{/search} (\textit{Search}),
\texttt{/evaluate} (\textit{Evaluate}) and \texttt{/abstract} (\textit{Abstract}). The \textit{Search} endpoint uses the arXiv API to query the papers by title,
and returns the basic meta data as JSON. The \textit{Evaluate} endpoint gets a paper title, arXiv ID, and some settings
such as the number of related papers to retrieve and the LLM choice. It then
uses the arXiv API to retrieve the paper contents, the Semantic Scholar API to 
retrieve full metadata from the paper and its citations, and recommended papers
to use as the base for the semantic similarity method. It uses Server-Side Events
(SSE) to to stream live updates to 
the frontend during evaluation. 
The full evaluation result, including the main paper, related work, extracted data and final report, is returned as JSON. It supports GPT-4o, GPT-4o mini
and Gemini 2.0 Flash as the LLMs for data extraction and evaluation, and uses 
SentenceTransformers~\citep{reimers-2019-sentence-bert} to generate vector
embeddings from the text\footnote{The encoder model used is \texttt{all-MiniLM-L6-v2}.}.
The \textit{Abstract} endpoint operates similarly but accepts only the title and
abstract as input. From these, it retrieves semantically related papers
and generates the novelty assessment. However, without access to the full paper
content, it cannot extract citations or construct the structured graph.
\section{GraphMind}

{GraphMind is a multi-source information display platform designed to help
users analyze scientific ideas by comparing them with relevant literature. It
presents a structured view of processed papers and supports novelty assessment
through interactive statistical insights. By integrating evidence extracted from
both the target paper and related works, the platform enables flexible retrieval
and visualization of relevant multi-source literature.} In what follows, we
first describe the search interface (Section~\ref{sec:search}), followed by 
the assessment results page (Section~\ref{sec:assessment-results}). More details of the implementation can be found in Appendix~\ref{sec:assess-pipeline}. 



\subsection{Search}\label{sec:search}

\begin{figure}[ht]
    \centering
    \includegraphics[width=1\columnwidth]{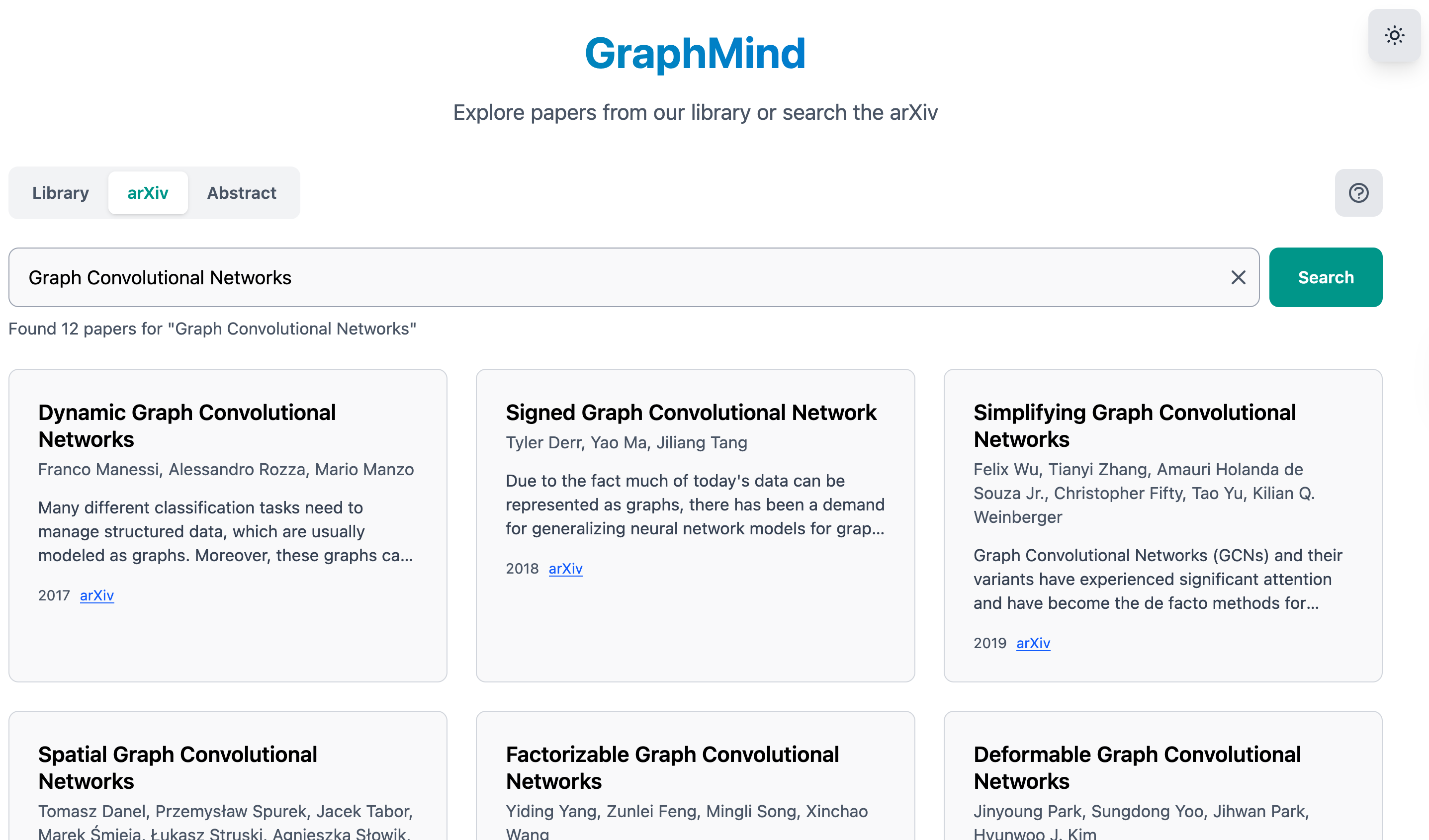}
    \caption{Search page showing the available tabs with the arXiv tab selected.}
    \label{fig:search-page}
\end{figure}

When the tool is launched, the user is directed to the
\textit{Search} page. On their first visit, a help message is displayed, offering
instructions on how to use the tool. After  dismissing this message, the user has three options to explore: the \textit{Library}, which contains a collection of pre-computed papers, the \textit{arXiv} search, and the \textit{Abstract}
evaluation. Figure~\ref{fig:search-page} shows the \textit{Search page} interface.


\paragraph{Library} This page shows a curated list of pre-computed papers available for immediate exploration. 
Each paper is presented as a card containing the title, abstract, and
publication venue. Users can click on a card to go to the \textit{Detail} page to see
the novelty assessment report.

\begin{figure}[h]
    \centering
    \includegraphics[width=\columnwidth]{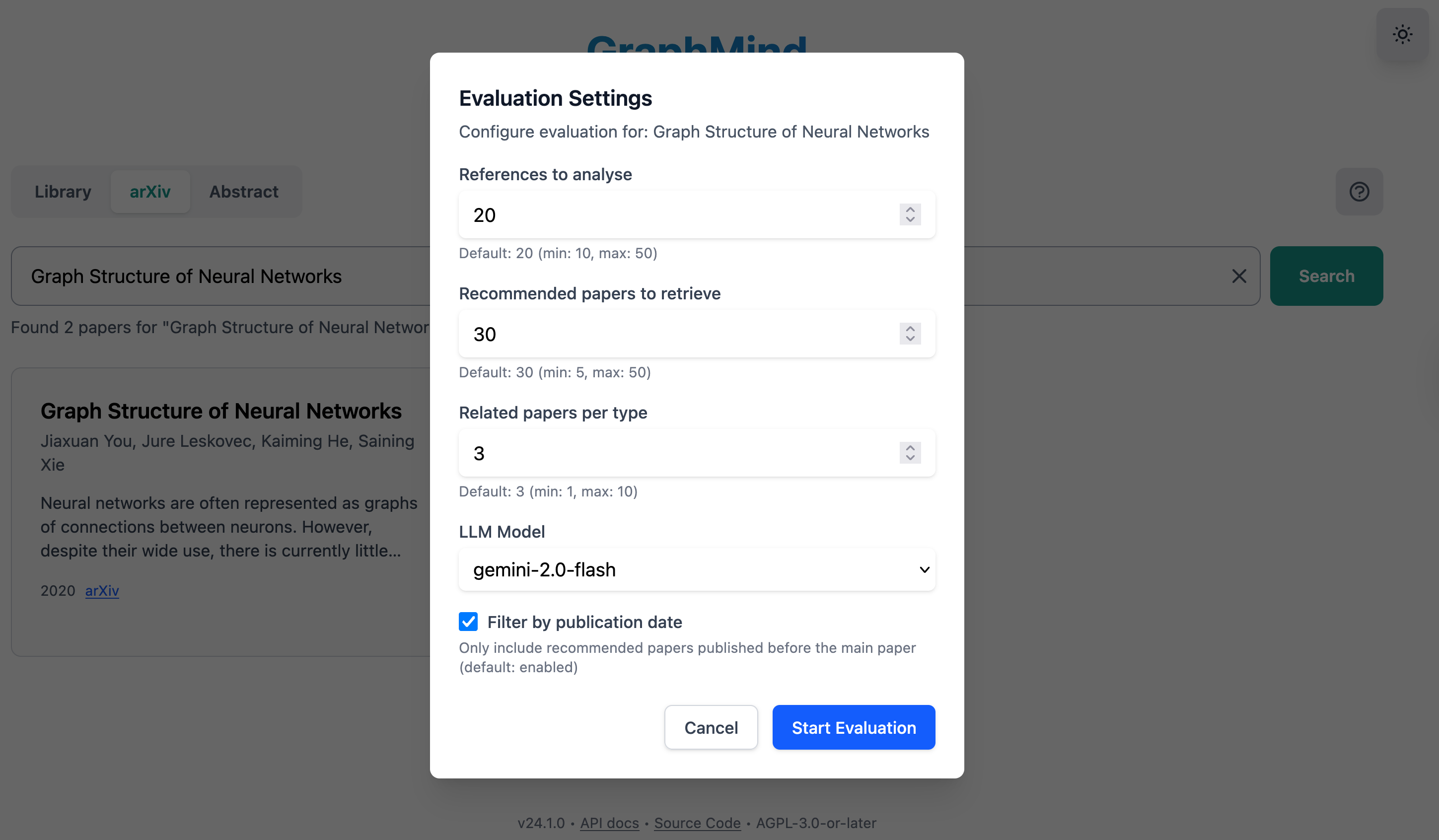}
    \caption{Customizable evaluation configuration panel.}
    \label{fig:configuration}
\end{figure}

\paragraph{arXiv} This page allows users to search for papers on arXiv by title,
using live data retrieval and dynamic analysis.
Results are presented 
similarly to those in the \textit{Library} tab. Upon selecting a paper, users are shown a configuration panel (see Figure~\ref{fig:configuration}) where they can customize evaluation settings, such as the number of citations to include, recommended papers to retrieve, related papers with which to
build the graph, and the LLM to use to extract the required information. They
can also choose whether to include all related papers or only those published before the main paper. 
Once confirmed, the assessment process starts (See Appendix~\ref{sec:assess-pipeline}).
During this process, a progress bar indicates the current steps being executed. Users 
can cancel at any time. When completed, they are redirected to the
\textit{Detail} page, which displays both micro- and macro-level novelty assessment results.

\paragraph{Abstract}

This tab allows users to directly enter a paper's title and abstract.
This enables the evaluation of papers not in the arXiv. The system 
extracts relevant information from the abstract and performs a macro-level novelty evaluation. Note that micro-level analysis is not possible due to lack of full arXiv metadata.


\subsection{Assessment Results}\label{sec:assessment-results} 


The \textbf{Detail} page provides comprehensive  paper analysis and novelty assessment results,
including \textbf{Metadata}, with basic paper information and predicted novelty
label, \textbf{Novelty Assessment}, with synthesized evaluation results and its
supporting evidence, \textbf{Paper Structured Graph} with micro-level elements
extracted from the paper, and \textbf{Related Papers}, with macro-level relevant papers
retrieved using both the citation network and semantic matches.
 


\begin{figure}[h]
    \centering
    \includegraphics[width=\columnwidth]{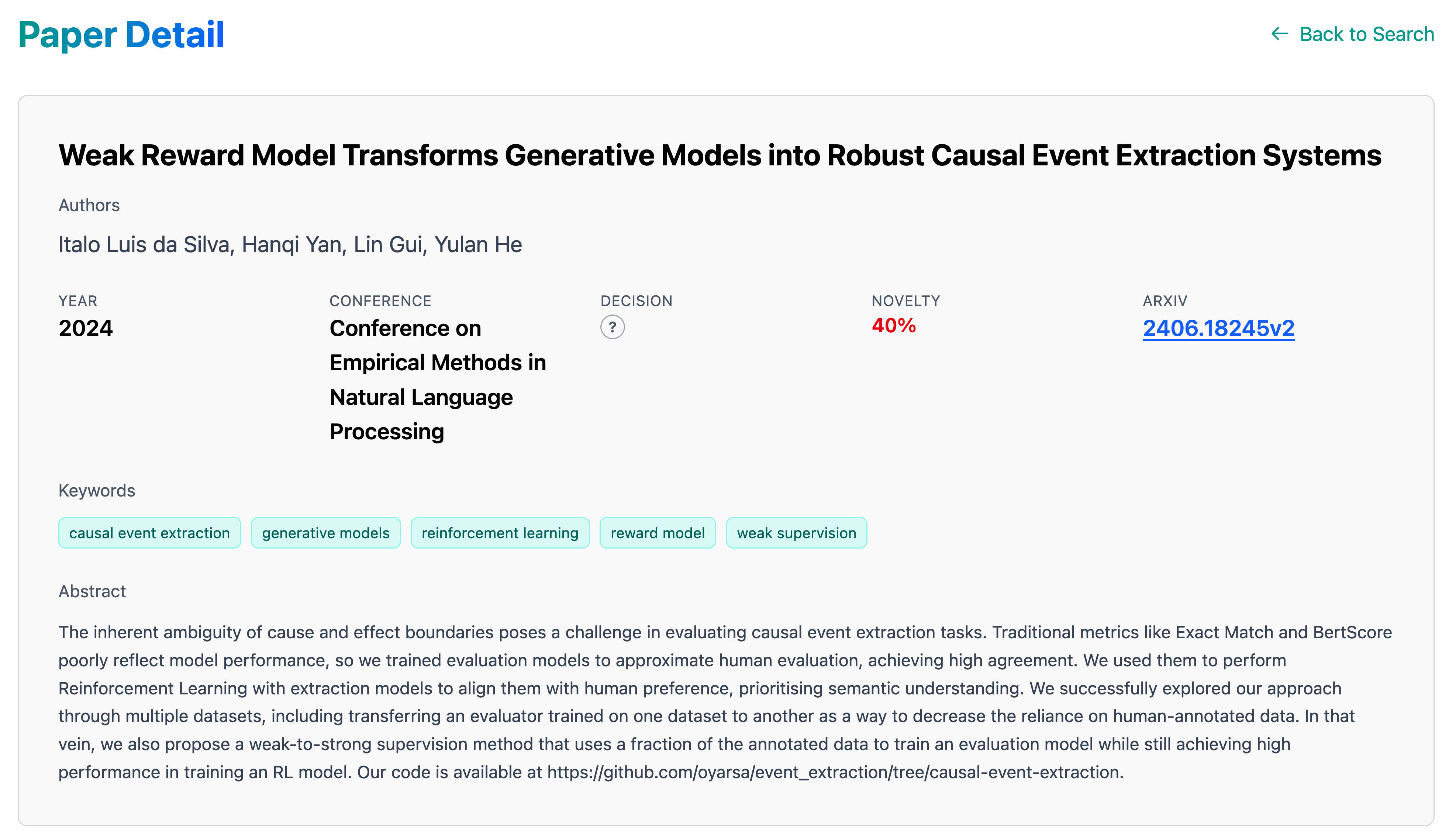}
    \caption{Metadata section on the Paper Detail page.}
    \label{fig:metadata-section}
\end{figure}

\paragraph{Metadata} Shows essential paper information: title, authors, publication year, conference
name, acceptance status (if available), arXiv link, extracted
keywords, and abstract. A novelty score, expressed as a percentage, is also provided. This score is obtained by prompting the evaluation
model for novelty assessment multiple times, and averaging the predictions. 
Figure~\ref{fig:metadata-section} shows the metadata section.



\begin{figure}[h]
    \centering
    \includegraphics[width=\columnwidth]{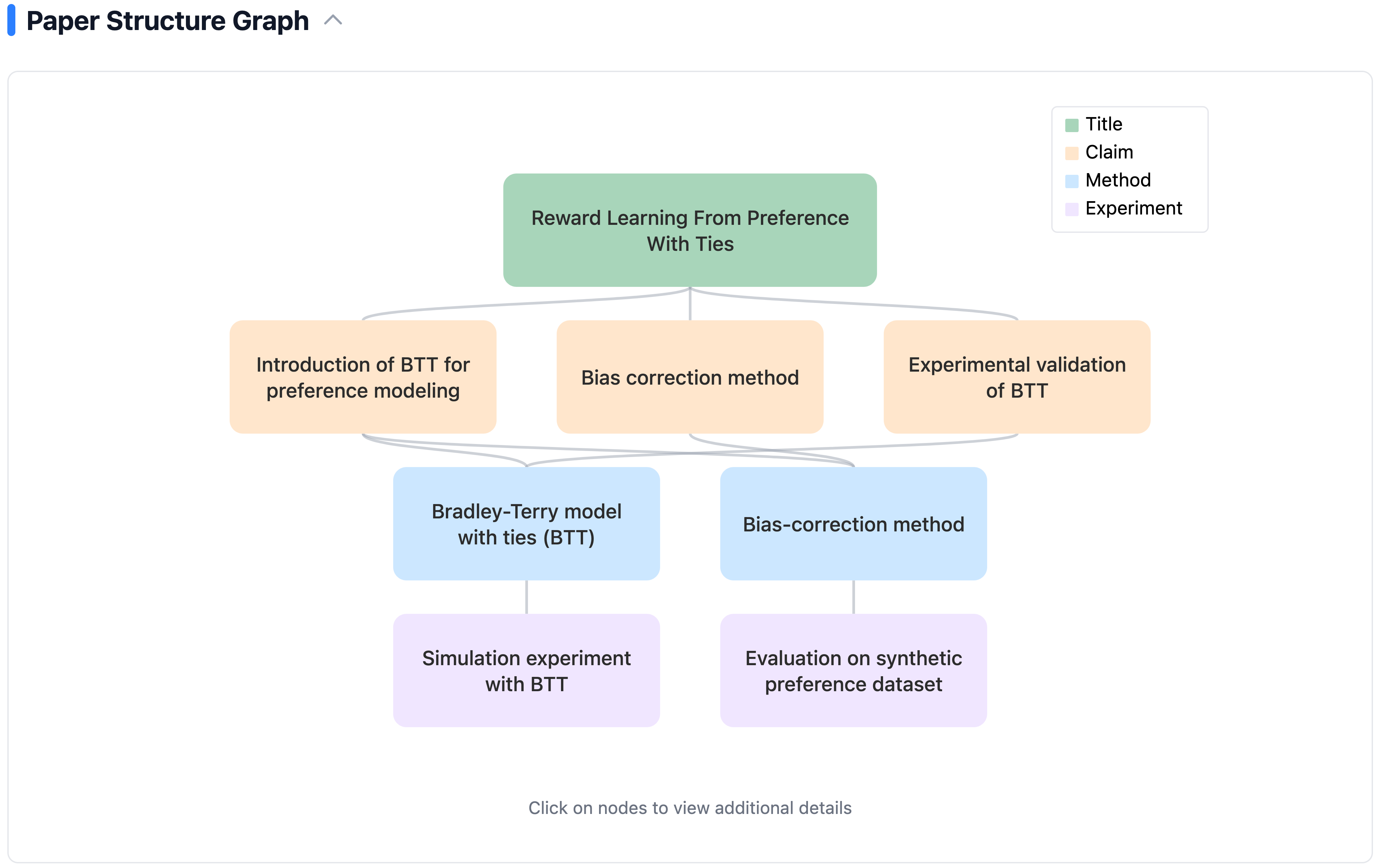}
    \caption{Paper structured graph.}
    \label{fig:graph-section}
\end{figure}

\paragraph{Paper Structured Graph} Presents the micro-level elements of the
paper extracted from the full paper content. It included the core claims, the methods used to validate those claims, and the experiments conducted as evidence
for the methods. These elements are interlinked, illustrating how claims are substantiated by 
methods, and how those methods are, in turn, validated through experiments. This structure helps users understand the key
aspects of the paper, how they support each other and how they contribute to the
overall argument. Users can interact with each node in the graph to view the corresponding segments extracted directly from the paper. 
Figure~\ref{fig:graph-section} shows the
structured graph.

\begin{figure}[h]
    \centering
    \includegraphics[width=\columnwidth]{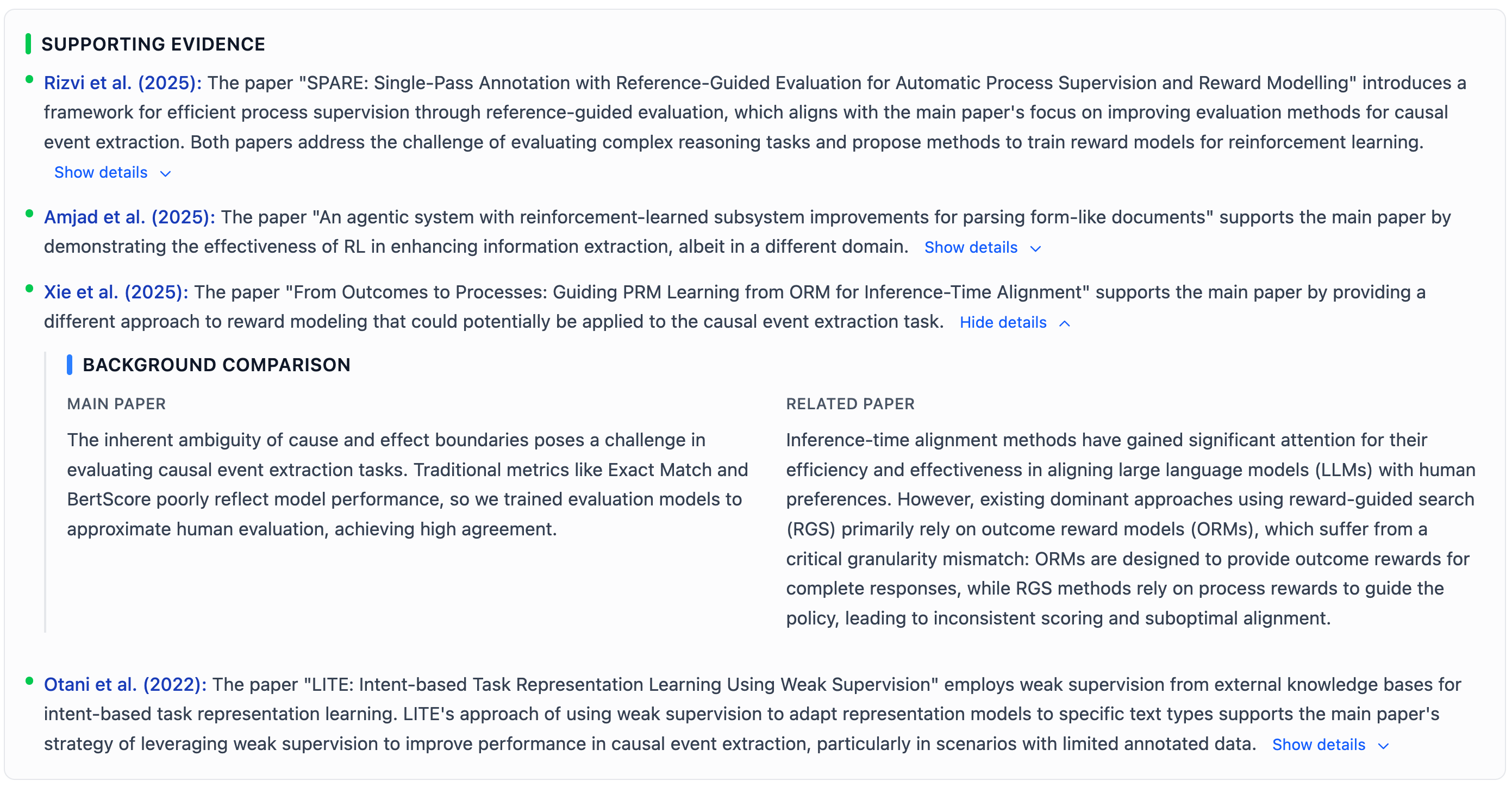}
    \caption{Novelty Assessment section showing supporting evidence comparison.}
    \label{fig:novelty-section}
\end{figure}

\paragraph{Novelty Assessment} Presents the results of novelty assessment using our proposed method.
It begins with a \textit{Result} summary that integrates the key findings
from both micro- and macro-level analyses, explaining how they collectively inform 
the final novelty score. Following the summary,  two sections provides deeper context: (1) \textbf{Supporting Evidence}: Highlights external papers that reinforce the novelty and soundness of the main paper’s approach. These examples show how the paper introduces new ideas or methods that are distinct from prior work. (2) \textbf{Contradictory Evidence}: Identifies related papers that challenge the originality or effectiveness of the proposed approach, pointing out overlaps or limitations.
Each evidence item links to a
related paper (see below) and highlights how that paper contributes to novelty 
assessment. 
Figure~\ref{fig:novelty-section} shows the
Novelty Assessment section with \textit{Supporting Evidence}, highlighting a
comparison of background information between the main and a related paper.



\paragraph{Related papers} The \textit{Related Papers} section displays the extracted information for each retrieved related paper.
This includes both the citations, categorized as either \emph{supporting} or \emph{contrasting} based on the context in which they are cited, 
and the semantically related papers, which are classified as either \emph{background} (works that inform or motivate the main paper) or \emph{ target} (works that address similar goals). For each related paper,
we show a semantic similarity score (both as a percentage and a visual scale), the original
abstract, and a relation-aware summary explaining the connection to the main paper. For cited papers, we also include
the citation contexts with their corresponding polarities (supporting or contrasting). For semantically related papers, we show either the extracted background content (for background papers) or the target content (for target papers).

\paragraph{Abstract evaluation}

For papers evaluated through the \textit{Abstract} input method, some features are unavailable due to the absence of full arXiv LaTeX content. Specifically, we cannot generate the \textbf{Paper Structured Graph} or include citation-based entries in the \textbf{Related Papers} section. 
However, we do provide semantically related
papers found based solely on the abstract content, and derive the novelty assessment from those relationships. 
This allows authors to position their in-progress work within the relevant literature, even in the absence of a full paper submission.

\section{Model Evaluation} To evaluate the performance of our methodology, we
conduct experiments using published papers from ICLR (2022-2025) and NeurIPS (2022-2024) conferences.
The papers were collected via the OpenReview
API\footnote{\url{https://docs.openreview.net/}}, and our full assessment
pipeline was applied to them. We use the median originality scores from peer reviews as our ground
truth. These scores range from 1 to 5 (see Appendix E for the complete scoring rubric), where we treat scores of 1-3 as "not
novel" and 4-5 as "novel". The predicted novelty 
labels are compared against the ground truth using standard classification metrics: precision, recall, F1 score, and accuracy. Our evaluation includes several LLMs: GPT-4o, Gemini 2.0
Flash, Qwen 2.5 7B~\citep{qwen2} and Llama 3.1 8B~\citep{Dubey2024TheLlama3}.

Table~\ref{tab:demo-perf} shows the performance of our method
against baselines. 
The \textit{Basic} baselines directly prompt a given LLM
using only the paper title and abstract, without our hierarchical graph or 
related papers. The \textit{Search} baselines are provided with our hierarchical graph. But they use the models' own web search capabilities 
to find relevant papers, not our curated retrieval pipeline. Note that both Qwen and Llama do not support search capabilities; therefore, we report results only for their Basic versions.

\begin{table}[h]  
    \centering
    \resizebox{0.99\columnwidth}{!}{%
    \begin{tabular}{@{}lcccc@{}}
        \toprule
        \textbf{Model} & \textbf{Precision} & \textbf{Recall} & \textbf{F1} & \textbf{Accuracy} \\
        \midrule
        Basic\textsubscript{GPT-4o}     & 0.6863 & \textbf{0.7000} & 0.6931 & 0.6900 \\
        Search\textsubscript{GPT-4o}    & 0.6667 & 0.6400 & 0.6531 & 0.6600 \\
        GraphMind\textsubscript{GPT-4o} & \textbf{0.7805} & 0.6400 & \textbf{0.7033} & \textbf{0.7300} \\
        \midrule
        Basic\textsubscript{Gemini}     & 0.5169 & \textbf{0.9200} & 0.6619 & 0.5300 \\
        Search\textsubscript{Gemini}    & 0.6667 & 0.7200 & 0.6923 & 0.6800 \\
        GraphMind\textsubscript{Gemini} & \textbf{0.7800} & 0.7222 & \textbf{0.7500} & \textbf{0.7400} \\
        \midrule
        Basic\textsubscript{Qwen}     & \textbf{0.6680} & 0.7371 & 0.7008 & 0.5500 \\
        GraphMind\textsubscript{Qwen} & 0.5946 & \textbf{0.8800} & \textbf{0.7097} & \textbf{0.5800} \\
        \midrule
        Basic\textsubscript{Llama}     & 0.5062 & \textbf{0.8200} & \textbf{0.6260} & 0.5100 \\
        GraphMind\textsubscript{Llama} & \textbf{0.5125} & 0.8000 & 0.6247 & \textbf{0.5200} \\
        \bottomrule    
    \end{tabular}
    }
    \caption{Evaluation result with different LLMs.}
    \label{tab:demo-perf}
\end{table}


We find that the \textit{Basic} baseline  struggles to accurately assess novelty, as it lacks sufficient context and relies solely on the LLM's internal knowledge. 
The \textit{Search} baseline retrieves papers less relevant than our related paper retrieval method,
which leads to lower performance with noisy data. Our method, \textit{GraphMind}, consistently outperforms both baselines by leveraging a structured representation of the paper and high-quality retrieved related works.

In addition to the baselines, we also select a few variations on our method
to show how each component contributes to the performance.
Table~\ref{tab:ablation-results} presents the results on our benchmark dataset
\texttt{SciNova} and PeerRead.

\begin{table}[H]
    \centering
    \resizebox{0.99\columnwidth}{!}{%
    \begin{tabular}{@{}lcccc@{}}
        \toprule
        \textbf{Variants} & \textbf{Precision} & \textbf{Recall} & \textbf{F1} & \textbf{Accuracy} \\
        \midrule
        \texttt{GraphMind} & 0.7800 & \textbf{0.7222} & \textbf{0.7500} & \textbf{0.7400} \\
        No citation & 0.7506 & 0.6704 & 0.7083 & 0.7200 \\
        No semantic & \textbf{0.8167} & 0.6772 & 0.7403 & 0.7200 \\
        No related & 0.8151 & 0.6703 & 0.7353 & 0.6900 \\
        No graph & 0.7217 & 0.6693 & 0.6942 & 0.6900 \\
        \bottomrule
    \end{tabular}
    }
    \caption{Ablation results with updated runs.}
    \label{tab:ablation-results}
\end{table}

We also evaluate the generated rationales using a Bradley-Terry tournament model ~\citep{bradley1952rank,chiang2023chatbot}. We employ an LLM (GPT-4o) as a judge to perform pairwise comparisons between 
the \textit{Basic} baseline, our \textit{GraphMind} method, and the original
human reviews. The comparisons are scored across the following five dimensions: 

\begin{tcolorbox}[title=Different aspects
of the generated rationales:, colback=gray!5, colframe=gray!40!black, breakable, fonttitle=\small, fontupper=\small]
{\textbf{Clarity}}: how easy is it to understand and to follow its
    ideas?

\vspace{0.8em}

{\textbf{Faithfulness}}: does the rationale justify the novelty
    label? For example, if the text is mostly positive, so should the label.

\vspace{0.8em}

{\textbf{Factuality}}: is the rationale is correct grounded in
    scientific facts from the target and related papers?

\vspace{0.8em}

{\textbf{Specificity}}: does the rationale cover information specific to the paper, or doe sit make overly generic statements? 

\vspace{0.8em}

{\textbf{Contributions}}: does the rationale effectively compare
    the target paper with the related papers? 

\end{tcolorbox}


Table~\ref{tab:demo-tournament} displays the results, showing that GraphMind outperfoms the baseline in general, and closely matches or exceeds human reviews in several aspects, particularly in \emph{Faithfulness}, \emph{Factuality}, and \emph{Specificity}. 


\begin{table}[h]
    \centering
    \resizebox{0.99\columnwidth}{!}{%
    \begin{tabular}{@{}lccccc@{}}
        \toprule
        \textbf{Model} & \textbf{Clarity} & \textbf{Faithful} & \textbf{Factuality} 
            & \textbf{Specificity} & \textbf{Contrib.} \\
        \midrule
        Human & \textbf{1547} & 1476 & 1470 & 1443 & \textbf{1584} \\
        Basic & 1520 & 1507 & 1386 & 1369 & 1430 \\
        GraphMind & 1520 & \textbf{1552} & \textbf{1609} & \textbf{1657} & 1540 \\
        \bottomrule
    \end{tabular}
    }
    \caption{Bradley-Terry ratings from automated pairwise tournament with GPT-4o as a judge.}
    \label{tab:demo-tournament}
\end{table}

The experimental results indicate that  \textit{GraphMind} is an effective system for assessing
novelty of scientific papers. It produces more accurate novelty labels compared to directly prompting LLMs, and generates rationales that are on par with, or even superior to, human-written reviews in terms of faithfulness, factual grounding, and specificity.
\section{Conclusion and Future Work}

\textbf{GraphMind} is an easy-to-use interactive web tool where users can
generate evaluation reports from scientific papers. It aims to assist peer
reviewers and other academic users in the task of assessing the novelty of a
paper using both its key elements and relationship with the scientific
literature. We achieve this by fetching the paper's full content from the arXiv, using
it to extract the key elements, and pairing the paper with relevant papers 
from the literature extracted from a related papers graph.

In the future, we will consider further expanding our related paper retrieval
to larger datasets and developing more refined ways of finding relevant papers.
This includes building our own database containing millions of scientific papers and using more advanced LLM-driven methods to
enhance the search process.  
We'll also consider incorporating interactive user feedback and evaluation on
domains beyond Machine Learning.

\section*{Acknowledgments}
This work was supported in part by the UK Engineering and Physical Sciences Research Council through a Turing AI Fellowship (grant no. EP/V020579/1, EP/V020579/2).

\bibliography{custom}
\appendix
\clearpage
\newpage
\section*{Appendix}
\setcounter{table}{0}
\renewcommand{\thetable}{A\arabic{table}}
\setcounter{figure}{0}
\renewcommand{\thefigure}{A\arabic{figure}}

\section{Evaluation dataset}

Table~\ref{tab:demo-distribution} shows the distribution of novelty labels
in our dataset. 

\begin{table}[!htbp]
    \centering
    \resizebox{0.89\columnwidth}{!}{%
    \begin{tabular}{@{}lrrrr@{}}
        \toprule
        \textbf{Year} & \textbf{Count} & \textbf{Count \%} & \textbf{Novel} & \textbf{Novel \%} \\
        \midrule
        2022 & 534 & 17.4\% & 450 & 84.3\% \\
        2023 & 688 & 22.5\% & 555 & 80.7\% \\
        2024 & 929 & 30.3\% & 549 & 59.1\% \\
        2025 & 912 & 29.8\% & 456 & 50.0\% \\
        \midrule
        Total & 3063 & 100.0\% & 2010 & 65.6\% \\
        \bottomrule
    \end{tabular}
    }
    \caption{Distribution of scientific papers by year with novelty rates.}
    \label{tab:demo-distribution}
\end{table}

\section{More screenshots}

\begin{figure}[h]
    \centering
    \includegraphics[width=\columnwidth]{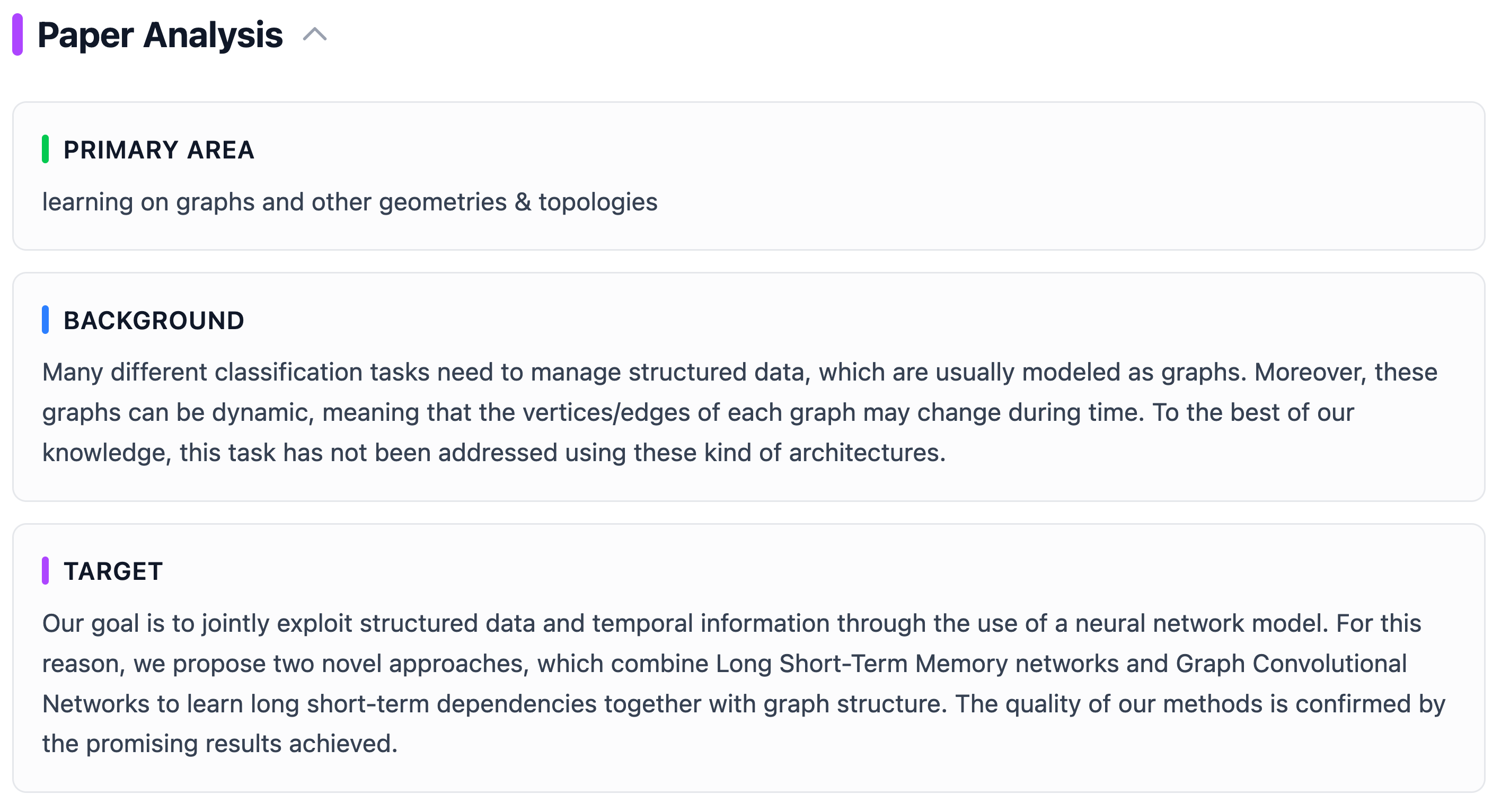}
    \caption{Paper Analysis section}
    \label{fig:analysis-section}
\end{figure}

\begin{figure}[h]
    \centering
    \includegraphics[width=\columnwidth]{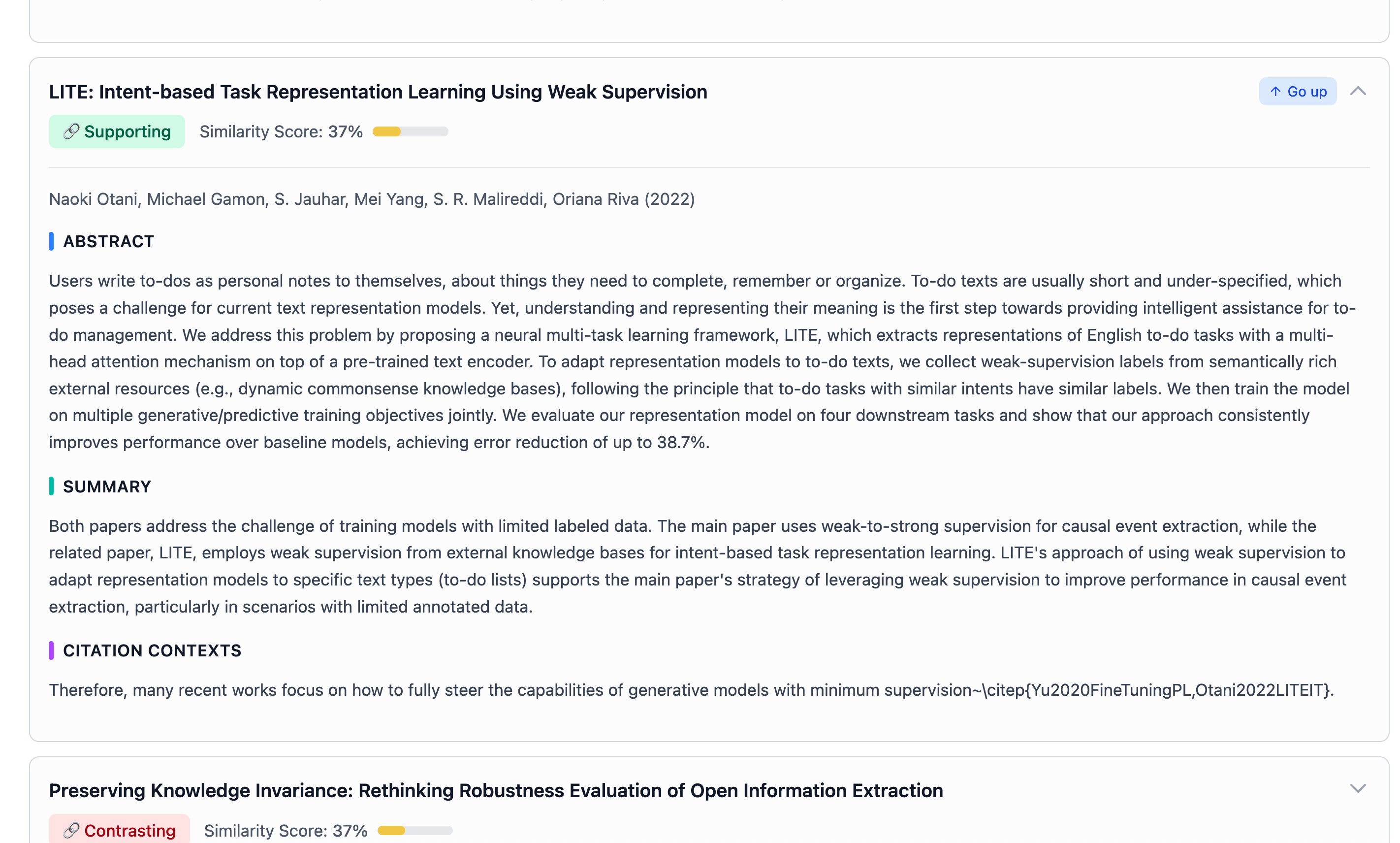}
    \caption{Related Papers section.}
    \label{fig:related-section}
\end{figure}

Figures~\ref{fig:analysis-section} and~\ref{fig:related-section} show the \textit{Paper Analysis} 
and \textit{Related Papers} sections, respectively.

\section{Assessment Pipeline}\label{sec:assess-pipeline}

Sections~\ref{sec:search} and~\ref{sec:assessment-results} show how our
tool looks like from the user's point of view. This section describes 
the pipeline that generates that information. This includes three
components:
\textbf{Graph extraction}, \textbf{Related papers retrieval} and the
final \textbf{Novelty assessment generation}. This section describes the steps
required for each one.

\paragraph{Graph extraction} We extract the graph from the full paper content
from the arXiv. First, we take the LaTeX code, extract citations from the
bibliography and convert the content to Markdown using
Pandoc\footnote{\url{https://pandoc.org/}}. For each citation, we identify all
sentences in the main text where it appears as citation contexts. An LLM then
extracts key components from the Markdown: claims, methods, experiments, and
their interconnections. For each component, we also extract supporting excerpts
from the paper.

\paragraph{Related Papers} We retrieve two types of papers: citations and
semantically related. Citations are extracted from the parsed
bibliography and filtered by semantic similarity to retain only the most
relevant ones. We then use an LLM to classify each citation context as positive
or negative. Citations are classified as supporting when contexts are mostly
positive, and as contrasting when contexts are mostly negative.

To find semantic neighbours, we first retrieve recommended papers via the
Semantic Scholar API. We then use an LLM to extract targets and backgrounds from
their abstracts, calculate semantic similarity with the main paper's targets and
backgrounds, and select the most relevant matches.

\paragraph{Novelty Assessment} We combine the paper graph and related papers to
generate the final novelty assessment. First, we convert the paper graph to text
using topological sorting to create a linear chain of nodes, then transform each
node into a paragraph using its label and supporting text. Second, we compile
the related papers into an evidence list, converting each paper into a paragraph
containing its title, relation type, and summary. Finally, we provide both
components as input to an evaluation LLM, which generates a novelty label and
structured rationale with result summary, supporting evidence, and contradictory
evidence

\section{Cost and Time}

Table~\ref{tab:time-cost} shows the average time and cost of performing
the full evaluation of an arXiv paper for each model available in the 
tool. Note that the cheapest and fastest model (Gemini 2.0 Flash) is also
the best performing (see~\ref{tab:demo-perf}.

\begin{table}[H]
    \centering
    \begin{tabular}{@{}lcc@{}}
        \toprule
        \textbf{Model} & \textbf{Time (s)} & \textbf{Cost (USD)} \\
        \midrule
        Gemini 2.0 Flash & 61.91 & 0.023213 \\
        GPT-4o & 75.06 & 0.477835 \\
        GPT-4o mini & 86.07 & 0.030429 \\
        \bottomrule
    \end{tabular}
    \caption{Time and cost of full novelty evaluation per model.}
    \label{tab:time-cost}
\end{table}

\section{Full novelty definition}

Our definition of novelty comes from the PeerRead paper~\citep{kang2018datasetpeerreviewspeerread}. We reproduce it here:

\begin{tcolorbox}[title={Novelty Definition},
    colback=gray!5!white,colframe=gray!75!black]
    
How original is the approach? Does this paper break new ground in topic,
methodology, or content? How exciting and innovative is the research it
describes?

Note that a paper could score high for originality even if the results do not
show a convincing benefit.

\begin{description}
    \item[5 = Surprising:] Significant new problem, technique, methodology, or
    insight -- no prior research has attempted something similar.
    
    \item[4 = Creative:] An intriguing problem, technique, or approach that is
    substantially different from previous research.
    
    \item[3 = Respectable:] A nice research contribution that represents a
    notable extension of prior approaches or methodologies.
    
    \item[2 = Pedestrian:] Obvious, or a minor improvement on familiar
    techniques.
    
    \item[1 = Significant portions] have actually been done before or done
    better.
\end{description}
\end{tcolorbox}

\end{document}